\documentclass[prb,aps,twocolumn]{revtex4-1}

\usepackage{color,amsmath,amssymb,amsfonts,graphicx,tabularx}

\usepackage[unicode=true,colorlinks=true]{hyperref}

\hypersetup{linkcolor=blue,citecolor=blue,urlcolor=blue}

\begin{document}

\title{Fractional quantum Hall states of bosons on cones}

\author{Ying-Hai Wu$^1$, Hong-Hao Tu$^{1,2}$, and G. J. Sreejith$^{3,4}$}

\affiliation{$^1$ Max-Planck-Institut f{\"u}r Quantenoptik, Hans-Kopfermann-Stra{\ss}e 1, 85748 Garching, Germany \\
$^2$ Physics Department, Arnold Sommerfeld Center for Theoretical Physics and Center for NanoScience, Ludwig-Maximilians-Universit{\"a}t M{\"u}nchen, 80333 M{\"u}nchen, Germany \\
$^3$ Indian Insitute for Science Education and Research, Pune 411008, India \\
$^4$ Max-Planck-Institut f{\"u}r Physik komplexer Systeme, 01187 Dresden, Germany
}

\date{\today}

\begin{abstract}
Motivated by a recent experiment which synthesizes Landau levels for photons on cones [Schine {\em et al.}, Nature 534, 671 (2016)], and more generally the interest in understanding gravitational responses of quantum Hall states, we study fractional quantum Hall states of bosons on cones. A variety of trial wave functions for conical systems are constructed and compared with exact diagonalization results. The tip of a cone is a localized geometrical defect with singular curvature which can modify the density profiles of quantum Hall states. The density profiles on cones can be used to extract some universal information about quantum Hall states. The values of certain quantities are computed numerically using the density profiles of some quantum Hall states and they agree with analytical predictions.
\end{abstract}

\maketitle

\section{Introduction}

The quantum Hall states observed in two-dimensional electron gases \cite{Klitzing1980,Tsui1982} are paradigmatic examples of topological phases of matter. Their defining characteristics are quantized Hall conductance and exponentially suppressed longitudinal conductance, which are largely independent of microscopic details and reflect the topological robustness of these states. In addition to the electromagnetic responses, the quantum Hall states also exhibit non-trivial features with respect to deformations of the metric. These gravitational responses are less studied than the electromagnetic responses, but they have attracted much attention recently~\cite{WenXG1992-2,Avron1995,Levay1995,Read2009,Read2011,Hoyos2012,Can2014,ChoGY2014,Abanov2014,YouY2014,Gromov2015,HuangB2015}. 

The gravitational responses contain universal coefficients that remain invariant under small perturbations. One can compute them by adiabatic variation of wave function on torus or using the Chern-Simons field theory. For example, a quantity called Hall viscosity can be defined to characterize the response to homogeneous change of spacetime metric \cite{Avron1995,Levay1995,Read2009}. For some quantum Hall states with identical electromagnetic properties, the Hall viscosity and other gravitational responses may help us to further distinguish them. It is in principle possible to measure the gravitational responses by creating local curvature on the surface where quantum Hall states reside. However, this will generally bring in some other undesirable perturbations and lead to non-universal phenomena. To this end, it has been proposed that one can study quantum Hall states on cones \cite{Biswas2016} where the spatial curvature is nonzero only at the tip and the quantum Hall states will only be perturbed in a small region around the tip without too much change of energetics.

Besides quantum Hall states of electrons in solid state systems, the possibility of realizing quantum Hall states of bosons has been actively pursued for some time. A variety of bosonic quantum Hall states have been studied theoretically in previous works \cite{Viefers2000,Cooper2001,Paredes2001,Regnault2003,ChangCC2005,Rezayi2005,Grass2012,FurukawaS2012,FurukawaS2013,WuYH2013,Regnault2013,Grass2014,Meyer2014,Geraedts2017}. The experimental realization of these states is very challenging because the bosons generally do not carry electric charges so do not couple to magnetic field as electrons do. There have been very exciting progresses in creating synthetic magnetic field for cold atoms and photons \cite{Gemelke2010,Aidelsburger2013,Miyake2013,Aidelsburger2015,Mittal2016,Schine2016}. In particular, a recent experiment demonstrated Landau levels of photons on cones \cite{Schine2016}. This paves the way towards directly measuring quantized gravitational responses of bosonic quantum Hall states.

This paper focuses on fractional quantum Hall (FQH) states of bosons on cones and is organized as follows. The main text studies one-component systems but some results of two-component systems are given in Appendix A. In Section II, we review the single-particle eigenstates on cones and define the many-body Hamiltonian for our systems. In Section III, we study several FQH states on cones using trial wave functions and exact diagonalizations. We conclude the paper with outlooks in Section IV. 

\section{Model}

\subsection{Single-particle Hamiltonian}

\begin{figure}
\includegraphics[width=0.48\textwidth]{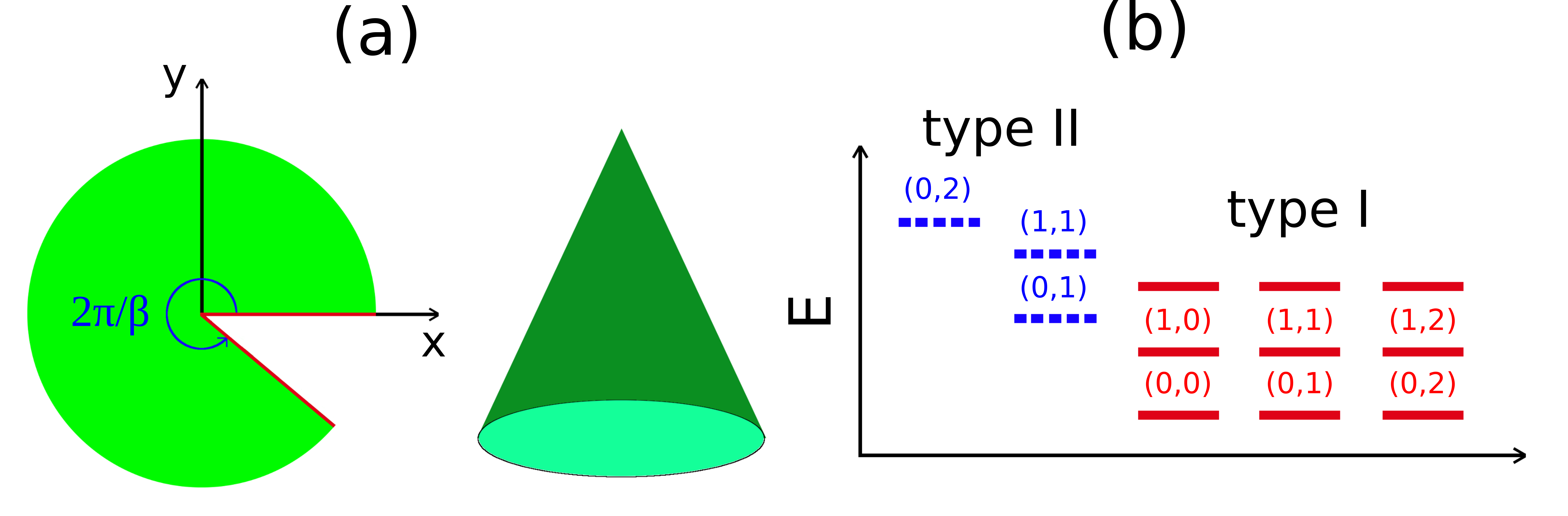}
\caption{(a) A cone can be bulit from a disk by removing a certain part and glue the resulting two edges together. (b) The Landau levels on a cone contain type I states Eq. (\ref{StateTypeI}) (red solid lines) and type II states Eq. (\ref{StateTypeII}) (blue dash lines). The quantum numbers for the states are displayed as $(s,m)$ above the lines.}
\label{Figure1}
\end{figure}

A cone can be built from a disk as shown in Fig. \ref{Figure1} (a): a section of the disk is removed (such that the remaining part spans an angle $2\pi/\beta$) and the resulting two edges are glued together. The single-particle Landau Hamiltonian is
\begin{eqnarray}
{\mathcal H}_0 = \frac{1}{2M} \left( {\mathbf p} - {\mathbf A} \right)^2
\end{eqnarray}
where the charge of the particles and the velocity of light are taken to be $1$ for simplicity, $M$ is an effective mass, and the gauge potential ${\mathbf A}=(-By/2,Bx/2)$ generates a uniform magnetic field through the cone. We define the magnetic length as $\ell=\sqrt{{\hbar}/B}$ (used as the unit of length in what follows) and the cyclotron frequency as $\omega_c=B/M$. The Cartesian coordinates on the cone are called $x$ and $y$, but the wave functions can be written more conveniently using the complex coordinate $z=(x+iy)/\ell$ with $\arg(z) \in [0,2\pi/\beta]$.

As shown in Fig.~\ref{Figure1} (b), there are two types of solutions to the single-particle problem \cite{Bueno2012}. The type I states
\begin{eqnarray}
\phi^{\rm I}_{s,m}({\mathbf r}) = {\mathcal N}_{s,m} z^{{\beta}m} L^{{\beta}m}_{s}\left(|z|^2/2\right) e^{-|z|^2/4}
\label{StateTypeI}
\end{eqnarray}
with $s=0,1,2,\cdots$ and $m=0,1,2,\cdots$ have eigenvalues $E^{\rm I}_{s,m}=(s+1/2)\hbar\omega_c$. The type II states
\begin{eqnarray}
\phi^{\rm II}_{s,m}({\mathbf r}) = {\mathcal N}_{s,m} z^{*{\beta}m} L^{{\beta}m}_{s}\left(|z|^2/2\right) e^{-|z|^2/4}
\label{StateTypeII}
\end{eqnarray}
with $s=0,1,2,\cdots$ and $m=1,2,\cdots$ have eigenvalues $E^{\rm II}_{s,m}=(s+{\beta}m+1/2)\hbar\omega_c$. The normalization factor is
\begin{eqnarray}
{\mathcal N}_{s,m} = (-1)^{s}\sqrt{\frac{{\beta}s!}{2{\pi}2^{{\beta}m}\Gamma(s+{\beta}m+1)}}
\end{eqnarray}
[$\Gamma(x)$ is the gamma function] and the Lagurrer polynomial is
\begin{eqnarray}
L^{{\beta}m}_{s}\left(|z|^2/2\right) = \sum^{s}_{k=0} (-1)^k \binom{s+{\beta}m}{s-k} \frac{1}{2^kk!} z^{*k} z^{k}
\end{eqnarray}
In the case of a flat disk with $\beta=1$, all the states with the same energy $(s+1/2)\hbar\omega_c$ form the conventional Landau level (LL) with orbital index $s$, which includes all the type I states $\phi^{\rm I}_{s,m}({\mathbf r})$ and the type II states $\phi^{\rm II}_{s_{\rm II},m_{\rm II}}({\mathbf r})$ satisfying $s_{\rm II}+m_{\rm II}=s$. For a general cone with $\beta\neq1$, these states will not have the same energy but we still group them together as a LL. In the subsequent discussions, we will use $\phi_{s,m}$ with no superscript ${\rm I}$ or ${\rm II}$ to denote both types of states. The states in the lowest LL (all the type I states with $s=0$) have very simple forms
\begin{eqnarray}
\phi^{\rm I}_{0,m}({\mathbf r}) = {\mathcal N}_{0,m} z^{{\beta}m} e^{-|z|^2/4}
\label{StateLLL}
\end{eqnarray}
which can be obtained from the states on a flat disk $({\sim}z^{m} e^{-|z|^2/4})$ by substituting $z$ with $z^{\beta}$ in the polynomial part and change the normalization factor accordingly. In the limit of $\beta\rightarrow\infty$, the cone resembles the thin cylinder or torus that has been studied before \cite{Seidel2005,Bergholtz2006,Bernevig2012-2}. However, the cone has an open end that extends to infinity so one can always go to a sufficiently large radius where the cone is not thin. This intuitive picture will be made more precise after we introduce the Hamiltonian.

\subsection{Many-body Hamiltonian}

The interaction between bosons is chosen to be the contact potential $V=4\pi\ell^2\sum_{i<j}\delta({\mathbf r}_i-{\mathbf r}_j)$. The prefactor $4\pi\ell^2$ is chosen such that the zeroth Haldane pseudopotential $P_{0}$ is $1$ \cite{Haldane1983}, which will be used as the unit of energy in all calculations. It is assumed that the bosons are confined to the lowest LL and there is no mixing with other LLs. The single-particle Hamiltonian projected to the lowest LL is a constant proportional to the total number of bosons and can be neglected. 

The number of bosons is denoted as $N_{b}$ and the creation (annihilation) operator for the state $\phi^{\rm I}_{0,m}({\mathbf r})$ is denoted as $C^\dagger_{m}$ ($C_{m}$). The second quantized form of the contact interaction is
\begin{eqnarray}
V &=& \sum_{\{m_i\}} V_{m_{1,2,3,4}} C^\dagger_{m_1} C^\dagger_{m_2} C_{m_4} C_{m_3}
\label{ManyHamilton1}
\end{eqnarray}
where the coefficient $V_{m_{1,2,3,4}}$ is
\begin{eqnarray}
\frac{\beta}{2} \left[ \prod^{4}_{i=1} \frac{2^{-{\beta}m_i}}{\Gamma({\beta}m_i+1)} \right]^{\frac{1}{2}} \Gamma\left[ \frac{\beta\sum_{i} m_{i}}{2} + 1 \right]
\end{eqnarray}
iff $m_{1}+m_{2}=m_{3}+m_{4}$ and zero otherwise. As $\beta$ increases, the terms with $m_{1}=m_{2}=m_{3}=m_{4}$ remain non-negligible but all other terms decay to zero in some different ways. For all the terms with $m_{1}=m_{3}$, $m_{2}=m_{4}$, the ones with $m_{1}=m_{2}{\pm}1$ have slower decay rate than the others. If one fixes the maximal $m_{i}$ and uses a sufficiently large $\beta$, $V$ can be approximated as the sum of its zeroth order part $\sum_{m} V^{0}_{m} n_{m} n_{m}$ and first order part $\sum_{m} V^{1}_{m} n_{m} n_{m+1}$. The $\beta$ value needed for this approximation to be valid increases with $m$, so the region closer to the cone tip will reach the thin cylinder/torus limit before the region far from the cone tip does.

The many-body Hamiltonian (\ref{ManyHamilton1}) is rotationally symmetric about the cone axis, so the total angular momentum $L_{z}=\sum^{N_{b}}_{i=1} m_{i}$ of a system is conserved. In experimental systems, an FQH state can form within a certain area only if the particles are confined by an external potential from escaping. To model this rotationally symmetric confinement, we assume that the total energy of a system has an additional term proportional to $L_{z}$ but this part will not be included explicitly in most discussions. One challenge in numerical calculations is that the number of single-particle states is infinite. For a system with a fixed total angular momentum $L_{z}$, the Hilbert space is finite as the maximum possible single-particle angular momentum is also $L_{z}$. However, the Hilbert space dimension may still be too large for numerical studies and it is desirable to further reduce the dimension by choosing a cutoff $L_{\rm max}$ for $m_{i}$. To make sure that the cutoff does not strongly affect the final results, we test multiple different choices to prove that the energy eigenvalues have converged very well.

\section{Results}

In this section, we first discuss some general properties of FQH states on cones and then study the Laughlin, Jain, and Moore-Read states \cite{Laughlin1983,Jain1989-1,Moore1991} in detail. The density $\rho({\mathbf r})$ of a quantum Hall state at filling factor $\nu$ is related to the magnetic field via
\begin{eqnarray}
\rho({\mathbf r}) = \nu \frac{B({\mathbf r})}{\Phi_{0}}
\label{DensityMagnetic}
\end{eqnarray}
where $\Phi_{0}=hc/e$ is the flux quantum. One manifestation of gravitational responses is that $\rho({\mathbf r})$ also varies with spatial curvature as described by
\begin{eqnarray}
\rho({\mathbf r}) = \nu \frac{B({\mathbf r})}{\Phi_{0}} + {\nu}S \frac{K({\mathbf r})}{4\pi}
\label{DensityCurvature}
\end{eqnarray}
where $S$ is a constant to be defined below and $K({\mathbf r})$ is the Gaussian curvature \cite{WenXG1992-2,Can2014,ChoGY2014,Biswas2016,Can2016}. The cones are particularly suitable for revealing the second term because the curvature of a cone is singular at its tip and vanishes everywhere else. The integral of the curvature over an area enclosing the cone tip is
\begin{eqnarray}
\int K({\mathbf r}) \; dA = 2\pi ( 1-\beta^{-1} )
\end{eqnarray}
For a flat disk without curvature, $\rho({\mathbf r})=\nu/{2\pi}$ in the bulk of the system and decreases to zero at the edge. When the state is realized on a cone, analytical studies suggest that $\rho({\mathbf r})$ is only perturbed in a small region around the cone tip and stays at $\nu/{2\pi}$ in the region that is far from both the tip and the edge \cite{WenXG1992-2,Can2016}. One can integreate over a certain area around the cone tip to obtain
\begin{eqnarray}
\int \rho({\mathbf r}) dA = \nu \int \frac{B({\mathbf r})}{\Phi_{0}} d A + \frac{{\nu}S}{2} ( 1-\beta^{-1} )
\label{DensityCone}
\end{eqnarray}
The second term means that there is a constant excessive charge accumulated around the cone tip. This phenomenon has been experimentally verified for integer quantum Hall states \cite{Schine2016}. A simple way to numerically compute the quantity $S$ in Eq. \ref{DensityCurvature} is to study the state on sphere \cite{Haldane1983}. The sphere has uniform curvature satisfying the Gauss-Bonnet theorem $\int K({\mathbf r}) \; dA =4\pi$, so we integrate over the entire sphere and obtain
\begin{eqnarray}
N_{b} = \nu N_{\phi} + {\nu}S
\end{eqnarray}
The quantity $S$ is called shift of the state and can be read directly from its wave function on sphere, which allows us to predict the excessive charge accumulated around the cone tip.

One can compute the density profiles on cones with different $\beta$ to verify Eq. \ref{DensityCone}. Because of the rotational symmetry of the cone, the density $\rho({\mathbf r})$ only depends on the radial distance to the cone tip and can be written as $\rho(r)$. For an eigenstate $|\Psi\rangle$ exressed in Fock space (e.g. those obtained in exact diagonalization), the density can be calculated directly as
\begin{eqnarray}
\rho({\mathbf r}) = \sum_{m} \left| \phi^{\rm I}_{0,m}(\mathbf r) \right|^2 \langle\Psi| C^\dagger_{m} C_{m} |\Psi\rangle.
\end{eqnarray}
For a real space wave function $\Psi({\mathbf r}_{1,2,\cdots,N_{b}})$, its density can be estimated using Monte Carlo methods as
\begin{eqnarray}
\rho(\mathbf{r}) &=& \left\langle \sum^{N_{b}}_{j=1} \delta({\mathbf r}-{\mathbf r}_{j}) \right\rangle \nonumber \\
&=& \frac{\sum^{N_{b}}_{j=1} \int \prod_{k} d{\mathbf r}_{k} \;\; \delta({\mathbf r}-{\mathbf r}_{j}) \left| \Psi({\mathbf r}_{1,2,\cdots,N_{b}}) \right|^{2} } { \int \prod_{k} d{\mathbf r}_{k} \;\; \left| \Psi({\mathbf r}_{1,2,\cdots,N_{b}}) \right|^{2} }
\label{FirstDensity}
\end{eqnarray}
It is expected that $\rho(r)$ shows a peak around the cone tip, takes the uniform value $\nu/{2\pi}$ in a large part of the cone, and decreases to zero at the edge. The excessive charge ${\Delta}n$ around the cone tip can be defined as
\begin{eqnarray}
{\Delta}n = \int_{r<r_{\rm max}} d{\mathbf r} \left[ \rho({\mathbf r}) - \frac{\nu}{2\pi} \right]
\end{eqnarray}
where the radius of integration $r_{\rm max}$ must be in the flat density region (i.e. far from both the tip and the edge). Ref. \onlinecite{Can2016} also defined higher moments of the density profile as
\begin{eqnarray}
{\chi}_{n} = \int_{r<r_{\rm max}} d{\mathbf r} \left[ \rho({\mathbf r}) - \frac{\nu}{2\pi} \right] \frac{r^{2n}}{2^n}
\end{eqnarray}
The excessive charge ($\Delta n = \chi_0$) and higher moments can be calculated using Monte Carlo as
\begin{equation}
{\chi}_{n} = \left\langle \sum^{N_{b}}_{i=1} \frac{|r_i|^{2n}}{2^{n}} \theta(r_{\rm max}-r_i) \right\rangle -\int_{r<r_{\rm max}} d {\mathbf r} \;\; \frac{\nu}{2\pi} \frac{r^{2n}}{2^n}
\end{equation}
where $\theta(r_{\rm max}-r_i)$ is the step function.

\subsection{Laughlin state}

The Laughlin state at $\nu=1/2$ on a flat disk is \cite{Laughlin1983}
\begin{eqnarray}
\Psi(\{z\}) = \prod_{j<k} (z_{j}-z_{k})^2
\label{LaughlinStateDisk}
\end{eqnarray}
For this and subsequent wave functions, we follow the convention of droping the ubiquitous Gaussian factor and only keep the polynomial part. The total angular momentum of this state is $L_{z}=N_{b}(N_{b}-1)$ and it is the highest density zero-energy eigenstate of the contact interaction $V$. By comparing the single-particle states on disk and cone, it is easy to see that the state should be
\begin{eqnarray}
\Psi(\{z\}) = \prod_{j<k} (z^{\beta}_{j}-z^{\beta}_{k})^2
\label{LaughlinStateCone}
\end{eqnarray}
on a cone. It is still an exact zero-energy eigenstate of $V$ because it vanishes as $|{\mathbf r}|^2$ when the distance ${\mathbf r}$ between any two bosons goes to zero \cite{Trugman1985}. It can be expanded in terms of symmetric monomials using the Jack polynomial method \cite{Bernevig2008}. The exact diagonalization results of several systems at $L_{z}=N_{b}(N_{b}-1)$ for various $\beta$ confirm that the ground states have zero energy and are identical to the Jack polynomial expansions.

\begin{figure}
\includegraphics[width=0.48\textwidth]{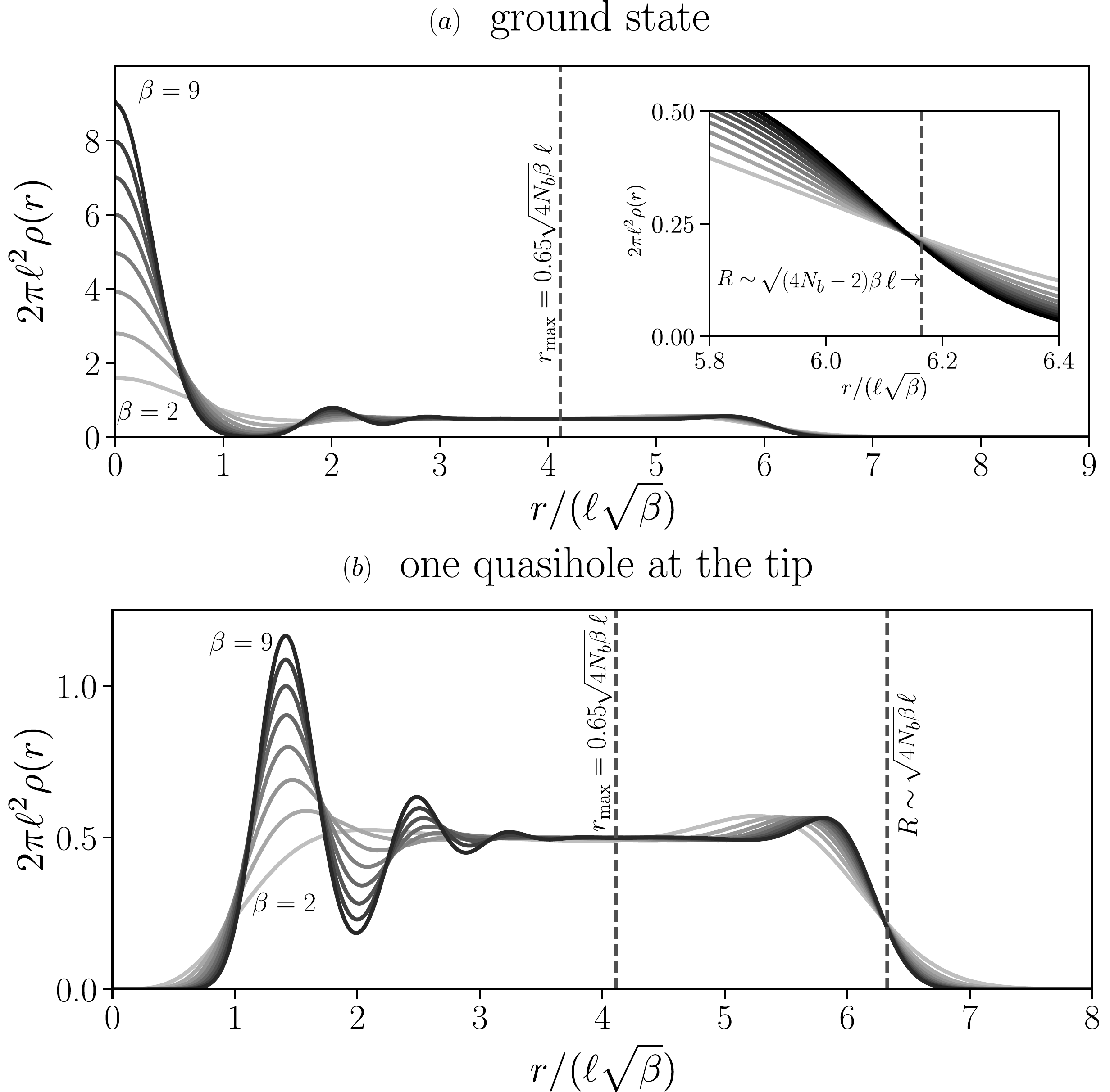}
\caption{The density profile $\rho(r)$ of the Laughlin state with $N_{b}=10$. (a) the ground state; (b) the state with one quasihole at the tip. The inset of panel (a) shows that radius of the droplet scales with $\sqrt{\beta}$ as we expect for a system with average density $1/(4\pi)$. $\ell$ is the magnetic length.}
\label{Figure2}
\end{figure}

\begin{figure}
\includegraphics[width=0.48\textwidth]{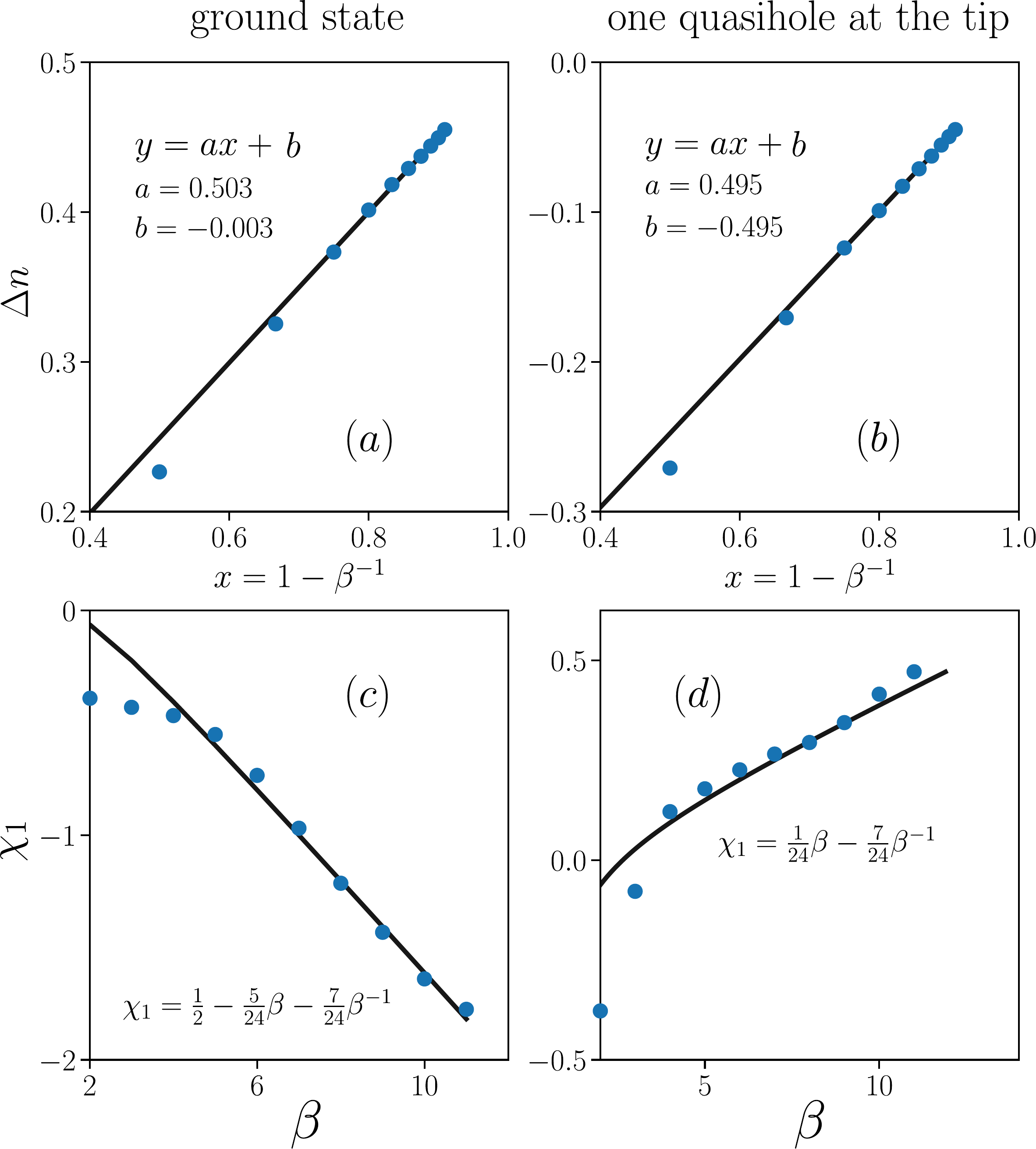}
\caption{The excessive charge ${\Delta}n$ and the first moment $\chi_{1}$ computed from Fig. \ref{Figure2}. (a) ${\Delta}n$ of the ground state; (b) ${\Delta}n$ of the state with one quasihole at the tip; (c) $\chi_{1}$ of the ground state; (d) $\chi_{1}$ of the state with one quasihole at the tip. The lines in panels (a) and (b) are linear fits using the data points at $\beta{\geq}4$. The lines in panels (c) and (d) are theoretical predictions (not fitting results using the data points).}
\label{Figure3}
\end{figure}

\begin{figure}
\includegraphics[width=0.48\textwidth]{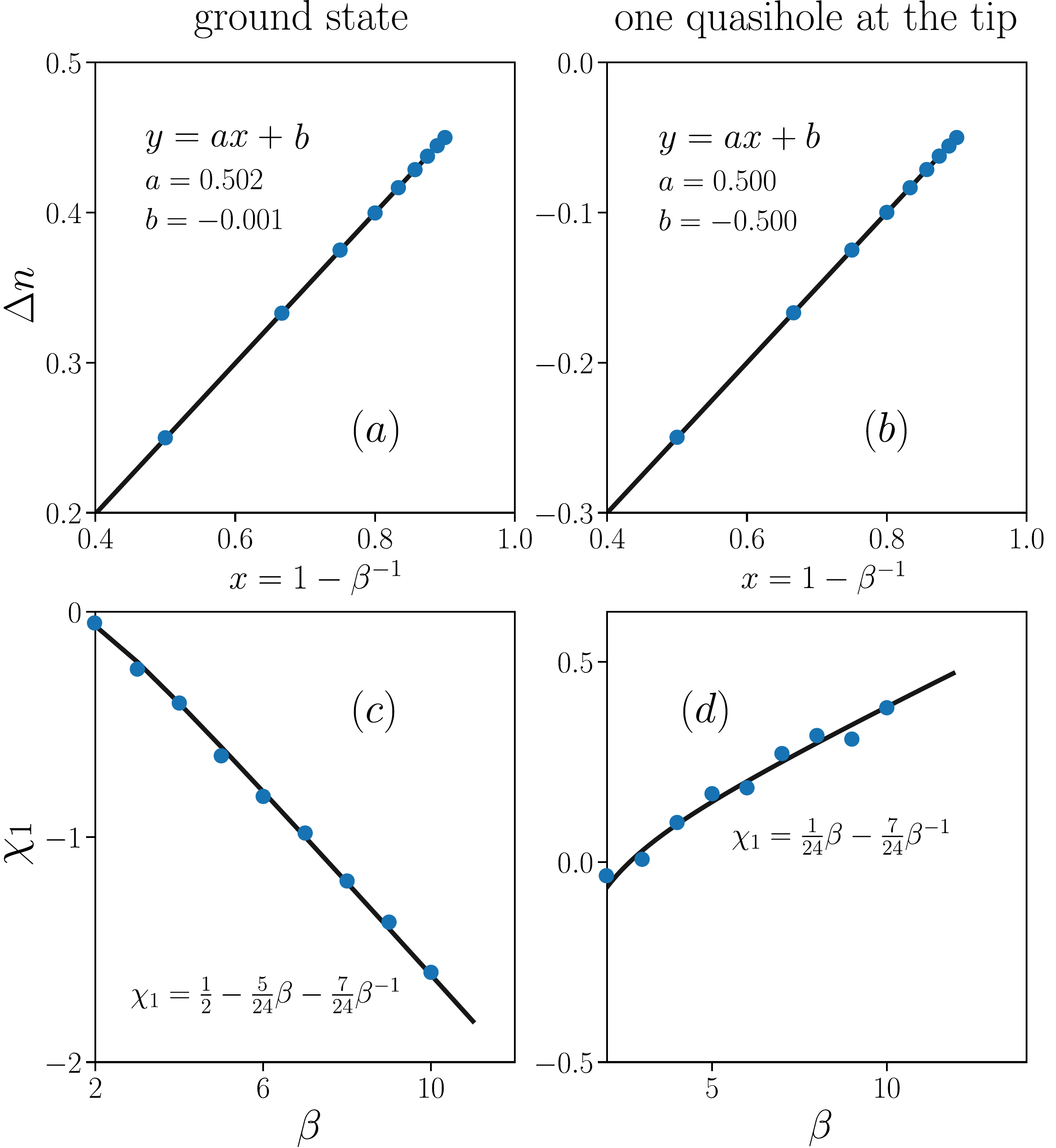}
\caption{The same quantities as in Fig. \ref{Figure3} but for the $N_{b}=40$ system. The data points at $\beta=2,3$ in panels (a) and (b) are also used in the linear fits and the agreement for $\chi_{1}$ is much better than Fig. \ref{Figure3}.}
\label{Figure4}
\end{figure}

The Laughlin $1/2$ state has $S=1$ so the excessive charge
\begin{eqnarray}
{\Delta}n = \frac{1}{2} (1-\beta^{-1})
\label{GroundCharge}
\end{eqnarray}
The first moment was predicted to be
\begin{eqnarray}
\chi_{1} = -\frac{5}{24}\beta - \frac{7}{24}\beta^{-1} + \frac{1}{2}
\end{eqnarray}
in Ref. \onlinecite{Can2016}. The number of bosons that can be reached in experimental systems is likely to be limited in the near future, so we would like to numerically check these relations using relatively small systems. The density profiles for many cases with $N_{b}{\leq}10$ have been computed and we find that these quantities cannot be extracted using $\beta{\leq}2$ because one cannot unambiguously identify a region with flat density $1/(4\pi)$. The large $\beta$ regime allows one to solve this issue but should be used with care. As we mentioned above, the Hamiltonian for a fixed $N_{b}$ approaches $\sum_{m} V^{0}_{m} n_{m} n_{m} + V^{1}_{m} n_{m} n_{m+1}$ with $V_m^1\ll V_m^0$ as $\beta$ increases. The ground state becomes the trivial product state $|101010\cdots\rangle$ in this limit. The energy gap in this angular momentum sector also decreases because it is of the order of $V^{1}_{m}$. One should make sure that the numerically extracted values are not due to the product state $|101010\cdots\rangle$ in a trivial way. Fig. \ref{Figure2} (a) shows our results in the $N_{b}=10$ system with $\beta\in[2,9]$, where the ground states are still sufficiently different from the product state. The excessive charge ${\Delta}n$ and the first moment $\chi_{1}$ computed from Fig. \ref{Figure2} (a) are shown in Fig. \ref{Figure3}. Eq. \ref{GroundCharge} is corroborated in Fig. \ref{Figure3} (a) where a linear fit of the data points at $\beta{\geq}4$ gives ${\Delta}n=0.503(1-\beta^{-1})-0.003$. The results at $\beta=2,3$ deviate from this fitting line because the flat density region is still not very flat. For the $\chi_{1}$ plot in Fig. \ref{Figure3} (c), the match between theoretical and numerical results is less impressive as $\chi_{1}$ is very senstitive to oscillations of $\rho(r)$. As we turn to the $N_{b}=40$ system in Fig. \ref{Figure4}, the linear fit of ${\Delta}n$ can be extended to $\beta=2,3$ and a much better agreement for $\chi_{1}$ is achieved.

In addition to the ground state, it is also useful to study quasihole states. The wave function for a system with one quasihole at the cone tip is
\begin{eqnarray}
\Psi(\{z\})_{\rm qh} = \prod_{j} z^{\beta}_{j} \prod_{j<k} (z^{\beta}_{j}-z^{\beta}_{k})^2
\end{eqnarray}
The density profile of this state can also reveal important information. The excessive charge is changed to
\begin{eqnarray}
{\Delta}n = \frac{1}{2} (1-\beta^{-1}) - \frac{1}{2}
\label{QuasiCharge}
\end{eqnarray}
because the quasihole carries $1/2$ charge and the first moment was predicted to be
\begin{eqnarray}
\chi_{1} = \frac{1}{24}\beta - \frac{7}{24}\beta^{-1}
\end{eqnarray}
in Ref. \onlinecite{Can2016}. Fig. \ref{Figure2} (b) shows our results in the $N_{b}=10$ system with $\beta\in[2,9]$. Eq. \ref{QuasiCharge} is corroborated in Fig. \ref{Figure3} (c) where a linear fit of the data points at $\beta{\geq}4$ gives ${\Delta}n=0.495(1-\beta^{-1})-0.495$. The accuracy of $\chi_{1}$ in Fig. \ref{Figure3} (d) is not very good but is greatly improved in Fig. \ref{Figure4} (d) using the $N_{b}=40$ system. 

\subsection{Jain states}

In addition to the $\nu=1/2$ Laughlin state, it has been found in previous works that FQH states also appear at larger filling factors. One important class is the Jain states at $\nu=n/(n+1)$ ($n\in{\mathbb Z}$) \cite{Regnault2003,ChangCC2005}. They can be understood using the composite fermion theory \cite{Jain1989-1} in which the bosons each absorb one magnetic flux and become composite fermions. It is sufficient to take the composite fermions as non-interacting objects in most cases, which form their effective LLs in an effective magnetic field. One can find good approximations to the low-lying states of the bosons by minimizing the effective cyclotron energy of the composite fermions in their effective LLs. In particular, if the composite fermions form an integer quantum Hall states (i.e., they fully occupy one or more effective LLs), the bosons are in an FQH state.

The filling factor of a finite-size system on a disk is not sharply defined because its density is not uniform but decreases at the edge. Nevertheless, one can roughly use the total angular momentum as a measure of the density. As we reduce the angular momentum from $N_{b}(N_{b}-1)$, the system will no longer possess zero-energy eigenstates but the composite fermion theory can help us to understand the physics in this regime. The general form of bosonic Jain states on a disk is
\begin{eqnarray}
\Psi(\{z\}) = {\mathcal P}_{\rm LLL} \Phi(\{z\}) \prod_{j<k} (z_{j}-z_{k})
\label{JainStateDisk}
\end{eqnarray}
where the Jastrow factor $\prod_{j<k} (z_{j}-z_{k})$ attaches one magnetic flux to each boson and $\Phi(\{z\})$ is a Slater determinant
\begin{eqnarray}
\det \left[
\begin{array}{cccc}
\phi_{s_{1},m_{1}}({\mathbf r}_{1}) & \phi_{s_{1},m_{1}}({\mathbf r}_{2}) & \cdots & \phi_{s_{1},m_{1}}({\mathbf r}_{N_{b}}) \\
\phi_{s_{2},m_{2}}({\mathbf r}_{1}) & \phi_{s_{2},m_{2}}({\mathbf r}_{2}) & \cdots & \phi_{s_{2},m_{2}}({\mathbf r}_{N_{b}}) \\
\cdots & \cdots & \cdots & \cdots \\
\phi_{s_{N_{b}},m_{N_{b}}}({\mathbf r}_{1}) & \phi_{s_{2},m_{2}}({\mathbf r}_{2}) & \cdots & \phi_{s_{N_{b}},m_{N_{b}}}({\mathbf r}_{N_{b}})
\end{array}
\right]
\end{eqnarray}
describing composite fermions in the single-particle states $\phi_{s_{1},m_{1}}$, $\phi_{s_{2},m_{2}}$, $\cdots$ and $\phi_{s_{N_{b}},m_{N_{b}}}$. The Laughlin state Eq. (\ref{LaughlinStateDisk}) is a special case of Eq. (\ref{JainStateDisk}) where the composite fermions only occupy their lowest effective LL. The energy of a many-body state can be quantified by the total effective cyclotron energy of the composite fermions. The states with the same effective cyclotron energy are expected to be quasi-degenerate and separated by energy gaps from others with different effective cyclotron energies. In reality, the quasi-degeneracy would only be clearly resolved if the number of states in that manifold is not too large otherwise the splittings between these states would be comparable to the gaps.

To construct Jain states on cones, we need to generalize the two parts of Eq. (\ref{JainStateDisk}). It is natural to guess that the Jastrow factor $\prod_{j<k} (z_{j}-z_{k})$ should be replaced by $\prod_{j<k} (z^{\beta}_{j}-z^{\beta}_{k})$. The counterpart of the Slater determinant $\Phi(\{z\})$ on a cone is less obvious. Are there also two types of states in the effective LLs of composite fermions? Do they have different energies when $\beta{\neq}1$? It will be demonstrated below that the Jain states on cones are
\begin{eqnarray}
\Psi(\{z\}) = {\mathcal P}_{\rm LLL} \Phi(\{z\}) \prod_{j<k} (z^{\beta}_{j}-z^{\beta}_{k})
\label{JainStateCone}
\end{eqnarray}
where $\Phi(\{z\})$ is constructed using the single-particle states on a cone with the same $\beta$ as the physical bosons. The relation between Eqs. (\ref{JainStateCone}) and (\ref{JainStateDisk}) are not as simple as in the Laughlin case. This is because one needs to use the single-particle states in higher LLs on the cones, which are not related to those on the disk in a simple way. The Jain states are given in real space but we can expand them in the Fock state basis, which helps us to compute their overlaps with the exact eigenstates. The main technical challenge in this process is the lowest LL projection (see Appendix B for more details).

\begin{figure}
\includegraphics[width=0.48\textwidth]{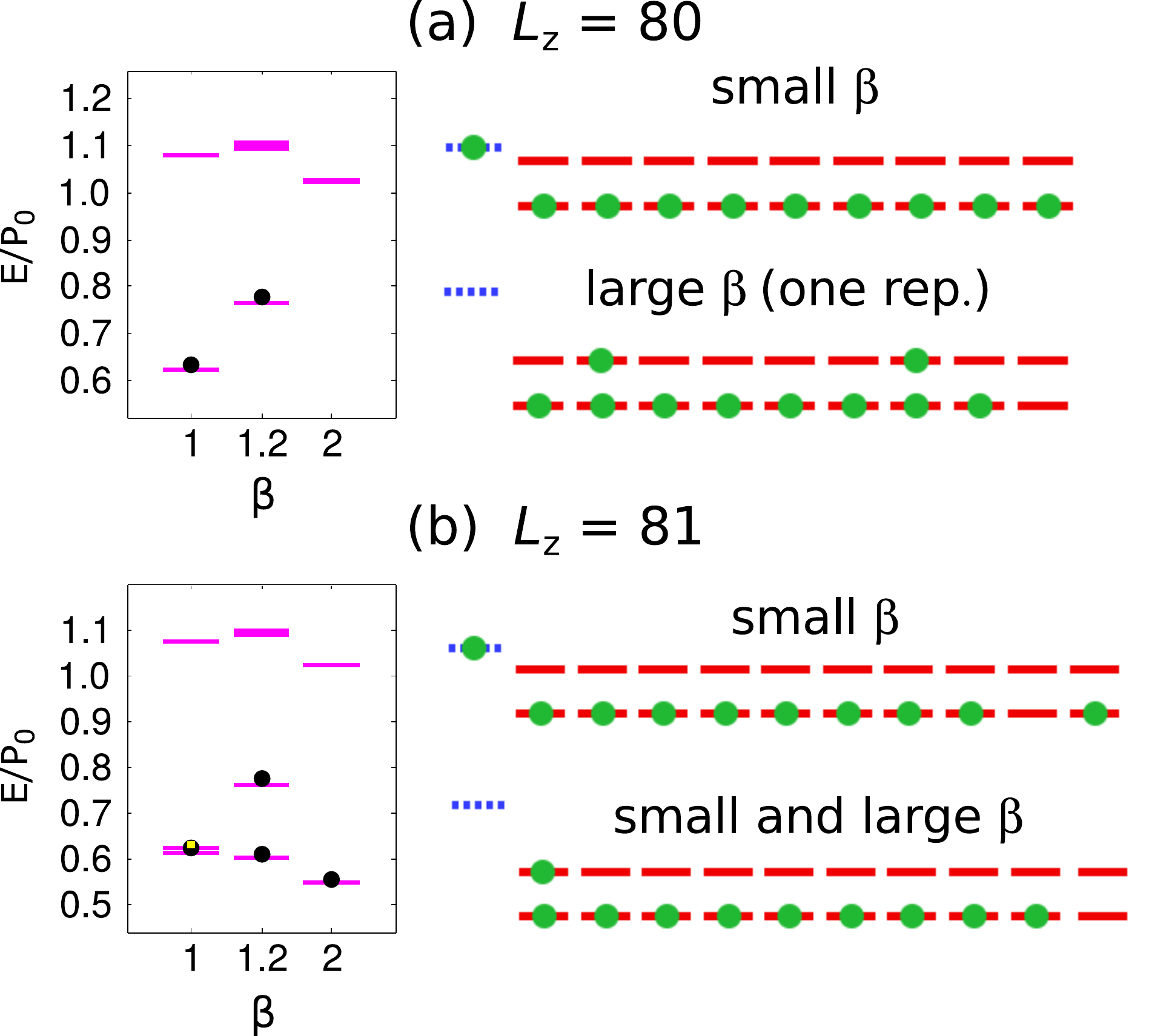}
\caption{The left parts show energy spectra of bosons on cones at $L_{z}=80$ and $81$ with $L_{\rm max}=25$ in units of $P_{0}$. The magenta lines are exact eigenstates and the black dots are composite fermion states. For $\beta=1.0$ at $L_{z}=81$, two states have very close energies so we use a yellow square in addition to a black dot. The right parts show the composite fermion configurations for the states. The red solid lines and blue dashed lines are defined as in Fig. \ref{Figure1}. The presence of a green dot means that this state is occupied. For $\beta=2.0$ at $L_{z}=80$, there are multiple configurations with the same effective cyclotron energy and we only give one representative.}
\label{Figure5}
\end{figure}

\begin{figure}
\includegraphics[width=0.48\textwidth]{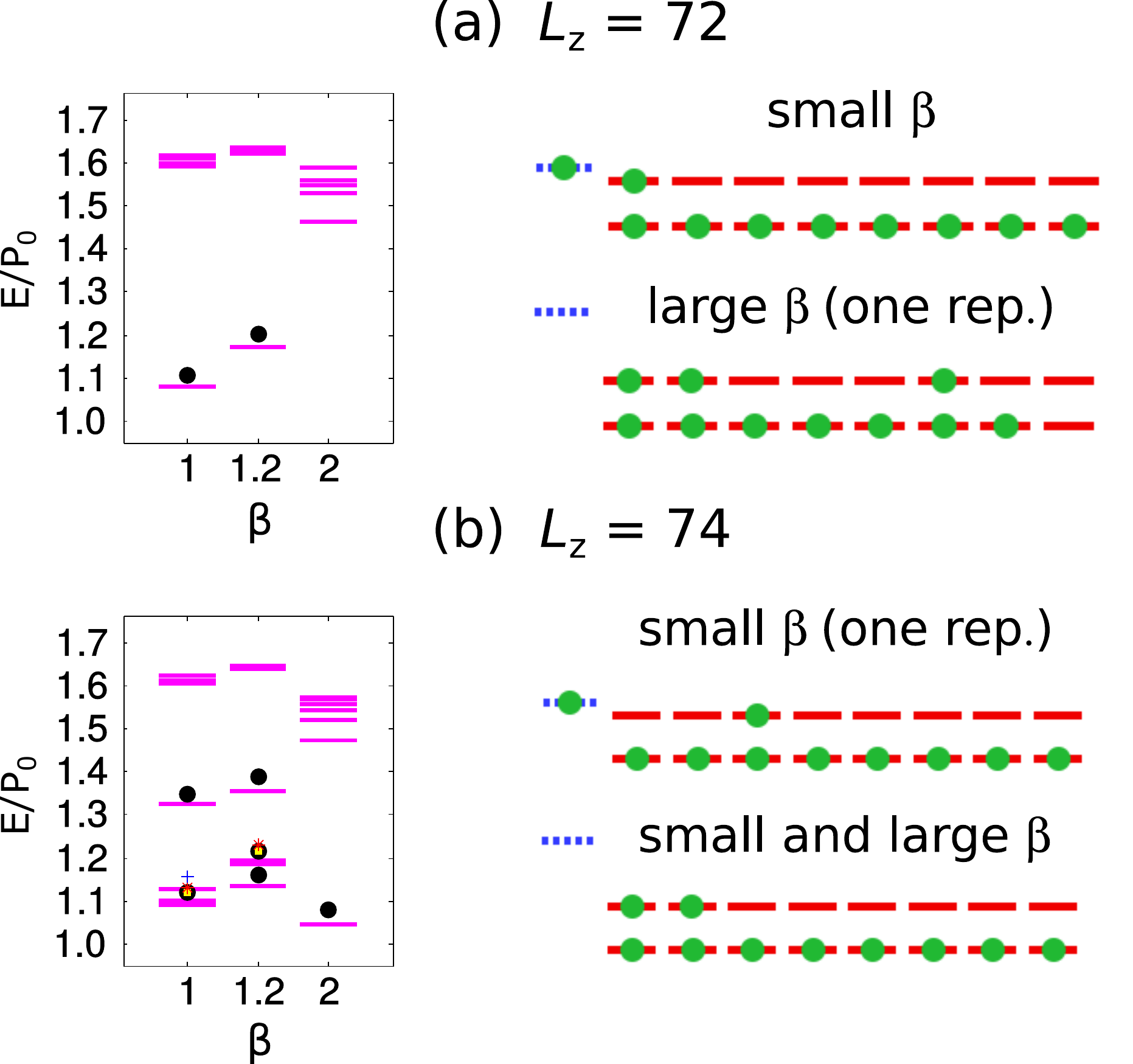}
\caption{The left parts show energy spectra of bosons on cones at $L_{z}=72$ and $74$ with $L_{\rm max}=25$ in units of $P_{0}$. The magenta lines are exact eigenstates and the black dots are composite fermion states. For $\beta=1.0$ and $\beta=1.2$ at $L_{z}=74$, several states have very close energies so we use a yellow square, a red star, and a blue cross in addition to a black dot. The right parts show the composite fermion configurations for the states. The red solid lines and blue dashed lines are defined as in Fig. \ref{Figure1}. The presence of a green dot means that this state is occupied. For $\beta=2.0$ at $L_{z}=72$ and $\beta=1.0$ and $1.2$ at $L_{z}=74$, there are multiple configurations with the same effective cyclotron energy and we only give one representative.}
\label{Figure6}
\end{figure}

We present the energy spectra for some $N_{b}=10$ systems in Figs. \ref{Figure5} and \ref{Figure6} with $L_{\rm max}=25$. The most important observation is that the number of quasi-degenerate states with the lowest effective cyclotron energy at a particular $L_{z}$ depends on $\beta$. This can be explained as due to varying energies of the type II states: some type II states may be occupied at small $\beta$ but they will not be favored when $\beta$ gets larger. Let us study the cases with $L_{z}=80$ and $72$ first. As shown in Figs. \ref{Figure5} (a) and \ref{Figure6} (a), exact diagonalizations find a state well-separated from others at $\beta=1.0$ and $1.2$. This can be explained using the composite fermion theory: there is a unique composite fermion configuration in which the state $\phi^{\rm II}_{0,1}$ is occupied. The gaps in the energy spectra vanish as $\beta$ increases to $2.0$ because the state $\phi^{\rm II}_{0,1}$ shifts to higher energy and is no longer occupied. It becomes energetically more favorable to place two composite fermions in the type I states of the $s=1$ effective LL, which results in multiple composite fermion configurations with the same effective cyclotron energy. For $L_{z}=81$ and $74$ shown in Figs. \ref{Figure5} (b) and \ref{Figure6} (b), we observe an opposite process: there are multiple low-energy states at $\beta=1.0$ and $1.2$ in which the state $\phi^{\rm II}_{0,1}$ is occupied but a unique ground state emerges as $\beta$ increases to $2.0$ where the system prefers not to occupy the state $\phi^{\rm II}_{0,1}$. The overlaps between the exact eigenstates and composite fermion states are given in Table \ref{Table1}.

\begin{table}
\centering
\begin{tabular}{ccccccc}
\hline\hline
Figure & ($L_{z}$,$\beta$) &        &        & overlap &        &       \\\hline
2(a)   & (80,1.0) & 0.9954 &        &        &        &        \\
       & (80,1.2) & 0.9925 &        &        &        &        \\
2(b)   & (81,1.0) & 0.9960 & 0.9954 &        &        &        \\
       & (81,1.2) & 0.9965 & 0.9924 &        &        &        \\
       & (81,2.0) & 0.9958 &        &        &        &        \\
3(a)   & (72,1.0) & 0.9813 &        &        &        &        \\
       & (72,1.2) & 0.9823 &        &        &        &        \\
3(b)   & (74,1.0) & 0.9809 & 0.9813 & 0.9801 & 0.9801 & 0.9851 \\
       & (74,1.2) & 0.9833 & 0.9006 & 0.9082 & 0.9673 & 0.9738 \\
       & (74,2.0) & 0.9798 &        &        &        &        \\
\hline\hline
\end{tabular}
\caption{The overlaps between exact eigenstates and composite fermion states in Figs. \ref{Figure5} and \ref{Figure6}. The numbers are ordered such that those with lower energy appear on the left.}
\label{Table1}
\end{table}

\subsection{Moore-Read state}

\begin{figure}
\includegraphics[width=0.48\textwidth]{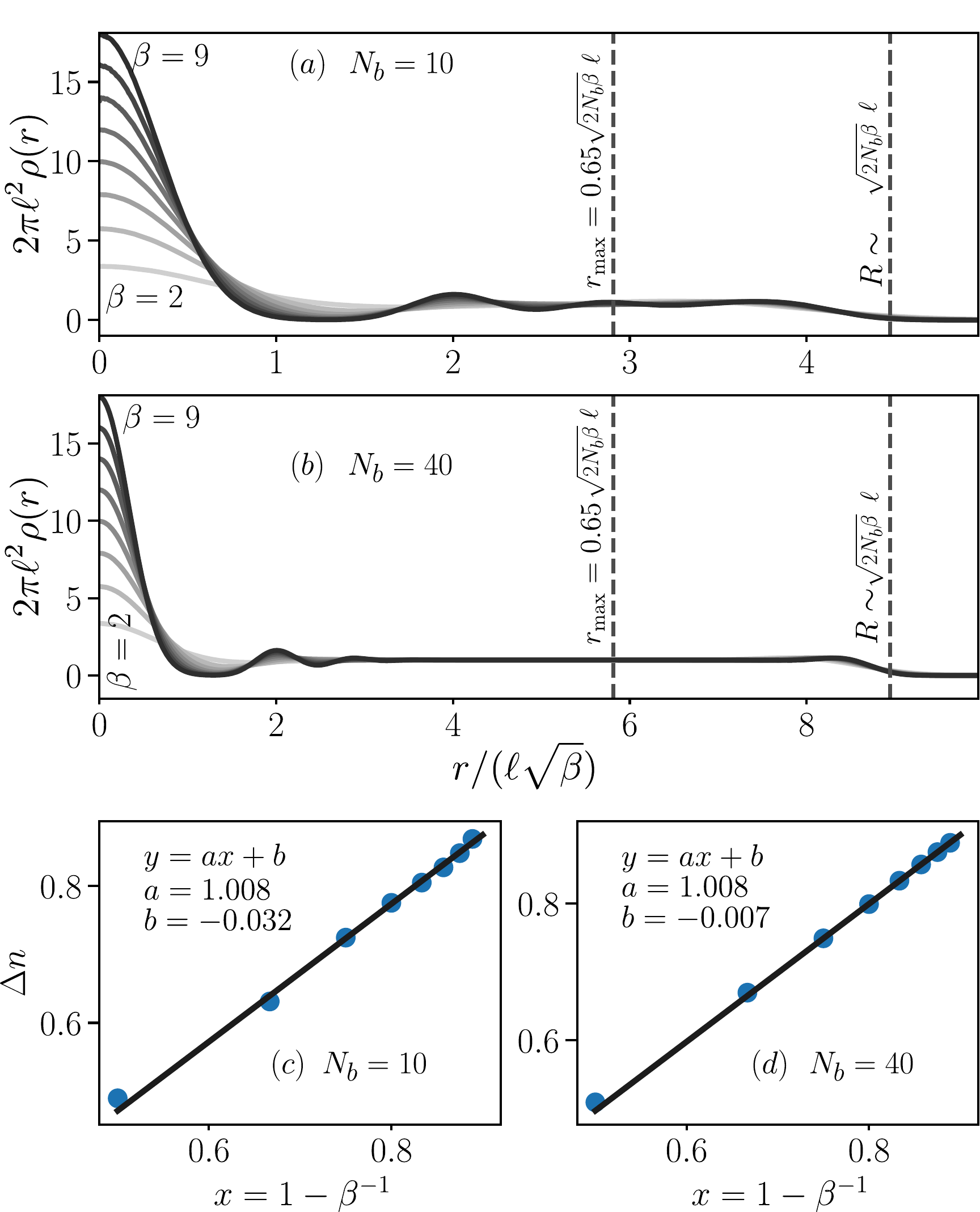}
\caption{The density profile $\rho(r)$ and the excessive charge ${\Delta}n$ of the Moore-Read state. (a) $\rho(r)$ of the $N_{b}=10$ system; (b) $\rho(r)$ of the $N_{b}=40$ system; (c) ${\Delta}n$ computed from panel (a); (d) ${\Delta}n$ computed from panel (b). The lines in panels (c) and (d) are linear fits where panel (c) uses the data points at $\beta{\geq}4$ and panel (d) uses all the data points.}
\label{Figure7}
\end{figure}

The Moore-Read state at $\nu=1$ on a flat disk is \cite{Moore1991}
\begin{eqnarray}
\Psi(\{z\}) = {\rm Pf}\left( \frac{1}{z_{j}-z_{k}} \right) \prod_{j<k} (z_{j}-z_{k})
\end{eqnarray}
where ${\rm Pf}$ is the Pfaffian of the matrix $1/(z_{j}-z_{k})$. This state has angular momentum $N_{b}(N_{b}-2)/2$, is the exact zero-energy eigenstate of the three-body contact interaction $\sum_{i<j<k}\delta({\mathbf r}_{i}-{\mathbf r}_{j})\delta({\mathbf r}_{j}-{\mathbf r}_{k})$, and can be expanded using the Jack polynomial method \cite{Bernevig2008}. In previous works, it was proposed that this state can be realized in systems with two-body contact interaction based on numerical results on sphere and torus~\cite{Regnault2003}. We have computed the ground states on disk for the $N_{b}=12$ system with several different cutoff $L_{\rm max}$. The overlap between the exact eigenstates and the Moore-Read state decreases rapidly as $L_{\rm max}$ gets larger (from $0.8276$ at $L_{\rm max}=10$ and $0.3444$ at $L_{\rm max}=15$). Based on the same argument used for the Laughlin state, the Moore-Read state on cones should be
\begin{eqnarray}
\Psi(\{z\}) = {\rm Pf}\left( \frac{1}{z^{\beta}_{j}-z^{\beta}_{k}} \right) \prod_{j<k} (z^{\beta}_{j}-z^{\beta}_{k})
\end{eqnarray}
It is still a zero-energy eigenstate which we confirm explicitly using exact diagonalization. This state is compared with the exact eigenstates of the $N_{b}=12$ system at several different $\beta$ and $L_{\rm max}$. The overlap also decreases rapidly as $L_{\rm max}$ gets larger (from $0.8492$ at $L_{\rm max}=10$ to $0.3905$ at $L_{\rm max}=15$ when $\beta=1.5$). These results suggest that it may be difficult to observe the Moore-Read state on cones with small $N_{b}$. The Moore-Read state has $S=1$ so the excessive charge
\begin{eqnarray}
{\Delta}n = 1-\beta^{-1}
\end{eqnarray}
Fig. \ref{Figure7} shows the density profiles and linear fits of ${\Delta}n$. It appears that the fitting coefficient is already quite accurate in the $N_{b}=10$ system [Fig. \ref{Figure7} (a) and (c)]. However, this is a coincidence because the result is unstable when we study the systems with $N_{b}=12{\sim}16$. This is due to oscillation of $\rho(r)$ in the central part of the cone and can be greatly suppressed as we go to $N_{b}{\geq}40$ [Fig. \ref{Figure7} (b) and (d)].

\section{Conclusion}

In conclusion, we have constructed trial wave functions for FQH states on cones, compared them with exact diagonalization results, and studied their gravitational responses to the singular curvature at the cone tip. The Laughlin state and Jain states are found to be accurate descriptions of bosons with contact interaction. However, the Moore-Read state is not a good approximation in many cases. The existence of two types of states in the single-particle spectrum is inherited by the composite fermions, which leads to dramatic consequences such as the (dis)appearance of unique ground states at certain angular momenta. Analytical predictions about the density profiles on the cones are substantiated by our numerical results. It would be very useful if one can develop an efficient projection method (something similar to the Jain-Kamilla procedure \cite{Jain1997}) to compute the density profiles of the Jain states and extract their gravitational responses.

\section*{Acknowledgement}

Exact diagonalization calculations are performed using the DiagHam package for which we are grateful to all the authors. This work was supproted by the DFG within the Cluster of Excellence NIM.

\begin{appendix}

\section{Two-Component Bosons}

\begin{figure}
\includegraphics[width=0.48\textwidth]{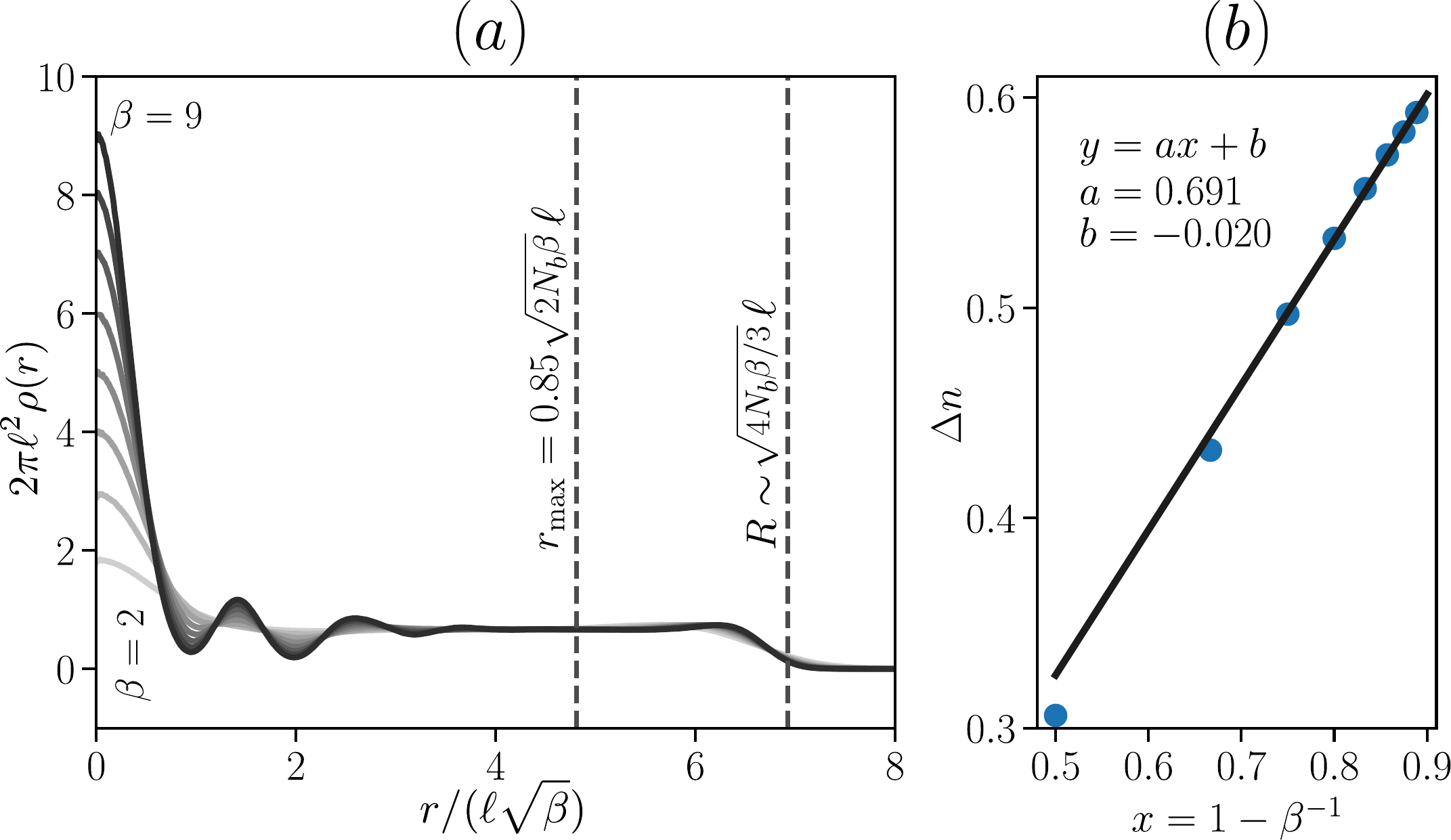}
\caption{The density profile $\rho(r)$ and the excessive charge ${\Delta}n$ of the Halperin 221 state with $N_{b}=16$. The line in panel (b) is a linear fit of the data points at $\beta{\geq}4$.}
\label{FigureA1}
\end{figure}

For two-component systems, the internal states will be labeled using $\sigma=\uparrow,\downarrow$. The number of bosons with internal state $\sigma$ is denoted as $N^{\sigma}_{b}$ ($N_{b}=N^{\uparrow}_{b}+N^{\downarrow}_{b}$ in this case) and the creation and annihilation operators are supplemented by a subscript $\sigma$. The bosons have contact interaction whose strength is independent of the internal states. The second quantized Hamiltonian is
\begin{eqnarray}
V &=& \sum_{\{m_i\}} V_{m_{1,2,3,4}} C^\dagger_{\sigma_1,m_1} C^\dagger_{\sigma_2,m_2} C_{\sigma_4,m_4} C_{\sigma_3,m_3}
\label{ManyHamilton2}
\end{eqnarray} 
with $V_{m_{1,2,3,4}}$ being the same as for one-component bosons. The zero-energy eigenstate in this case is the $\nu=2/3$ Halperin $221$ state \cite{Halperin1983}
\begin{eqnarray}
& \Psi(\{z^{\uparrow}\},\{z^{\downarrow}\}) = \prod_{j<k} (z^{\uparrow\beta}_{j}-z^{\uparrow\beta}_{k})^2 \nonumber \\
& \prod_{j<k} (z^{\downarrow\beta}_{j}-z^{\downarrow\beta}_{k})^2 \prod_{j,k} (z^{\uparrow\beta}_{j}-z^{\downarrow\beta}_{k})
\label{HalperinStateCone}
\end{eqnarray}
which we have checked explicitly in exact diagonalizations. This state has $S=2$ so the excessive charge
\begin{eqnarray}
{\Delta}n = \frac{2}{3} (1-\beta^{-1})
\end{eqnarray}
which is confirmed by our numerical results in Fig. \ref{FigureA1}.

\section{Lowest Landau Level Projection}

To expand Eq. (\ref{JainStateCone}) using symmetric monomials, we need to know the LLL projection
\begin{eqnarray}
{\mathcal P}_{\rm LLL} \phi_{s_{1},m_{1}} z^{{\beta}m_{2}}
\label{ProjectExpan}
\end{eqnarray}
for both type I and type II single-particles $\phi$. The Gaussian and normalization factors in $\phi$  are not important for these calculations so we will neglected them. The useful quantities are the coefficients $C^{{\rm I},m}_{s_{1},m_{1};m_{2}}$ and $C^{{\rm II},m}_{s_{1},m_{1};m_{2}}$ in
\begin{eqnarray}
&& {\mathcal P}_{\rm LLL} \;\;\; z^{{\beta}m_{1}} L^{{\beta}m_{1}}_{s_{1}}\left(|z|^2/2\right) z^{{\beta}m_{2}} \nonumber \\
= && \sum_{m} C^{{\rm I},m}_{s_{1},m_{1};m_{2}} z^{{\beta}m} \\
&& {\mathcal P}_{\rm LLL} \;\;\; z^{*{\beta}m_{1}} L^{{\beta}m_{1}}_{s_{1}}\left(|z|^2/2\right) z^{{\beta}m_{2}} \nonumber \\
= && \sum_{m} C^{{\rm II},m}_{s_{1},m_{1};m_{2}} z^{{\beta}m}
\end{eqnarray}
By multiplying $z^{{\beta}m}\exp(-|z|^2/2)$ on both sides and integrating over the entire cone, we find that the values are
\begin{eqnarray}
C^{{\rm I},m}_{s_{1},m_{1};m_{2}} &=& \delta_{m,m_{1}+m_{2}} \sum^{s_{1}}_{k=0} (-1)^{k} \frac{1}{k!} \binom{s_{1}+{\beta}m_{1}}{s_{1}-k} \nonumber \\
&\times& \frac{ \Gamma \left[ k+\beta(m_{1}+m_{2})+1 \right] }{ \Gamma \left[ \beta(m_{1}+m_{2})+1 \right] }\\
C^{{\rm II},m}_{s_{1},m_{1};m_{2}} &=& \delta_{m,m_{2}-m_{1}} \sum^{s_{1}}_{k=0} (-1)^{k} \frac{1}{k!} \binom{s_{1}+{\beta}m_{1}}{s_{1}-k} \nonumber \\
&\times& \frac{ 2^{{\beta}m_{1}} \Gamma \left[ k+{\beta}m_{2}+1 \right] } { \Gamma \left[ \beta(m_{2}-m_{1})+1 \right] }
\end{eqnarray}

\end{appendix}

\bibliography{../ReferCollect}

\end{document}